# A Neuron as a Signal Processing Device

Tao Hu[1,2], Zaid J. Towfic[1,3], Cengiz Pehlevan[1], Alex Genkin[1,4], and Dmitri B. Chklovskii[1]

[1] Janelia Farm Research Campus
Howard Hughes Medical Institute
Ashburn, VA 20147

[2] Texas A&M University
MS 3128 TAMUS
College Station, TX 77843

[3] Electrical Engineering Department
UCLA
Los Angeles, CA 90095

[4] AVG Consulting
Brooklyn, NY

*Abstract*— **A neuron is a basic physiological and computational unit of the brain. While much is known about the physiological properties of a neuron, its computational role is poorly understood. Here we propose to view a neuron as a signal processing device that represents the incoming streaming data matrix as a sparse vector of synaptic weights scaled by an outgoing sparse activity vector. Formally, a neuron minimizes a cost function comprising a cumulative squared representation error and regularization terms. We derive an online algorithm that minimizes such cost function by alternating between the minimization with respect to activity and with respect to synaptic weights. The steps of this algorithm reproduce well-known physiological properties of a neuron, such as weighted summation and leaky integration of synaptic inputs, as well as an Oja-like, but parameter-free, synaptic learning rule. Our theoretical framework makes several predictions, some of which can be verified by the existing data, others require further experiments. Such framework should allow modeling the function of neuronal circuits without necessarily measuring all the microscopic biophysical parameters, as well as facilitate the design of neuromorphic electronics.**

## I. Introduction

While the complexity of neuronal function is truly mind-boggling, several of its salient features can be summarized in the following diagram, Fig. 1A. A typical neuron receives spike trains from thousands of upstream neurons via synapses converting spikes into currents. The magnitude of each such current is called a synaptic weight and its shape indicates low-pass (LP) filtering, or leaky integration. Synaptic currents summate in the cell body, thus depolarizing the axonal membrane to generate a spike train that then is transmitted to thousands of downstream neurons. For the purposes of this paper, we quantify the incoming and the outgoing spike trains by a firing rate, which is the number of spikes per unit time. The firing rate is a nonlinear function of the summed current often approximated by a one-sided soft-threshold function. The firing rate of a neuron typically follows a nonGaussian distribution that is sparse (i.e. with a significant peak at zero) and heavy-tailed (i.e. decaying slower than a Gaussian). Finally, synaptic weights are continuously updated according to the correlation between the activity of presynaptic and postsynaptic neurons, known as Hebbian learning.

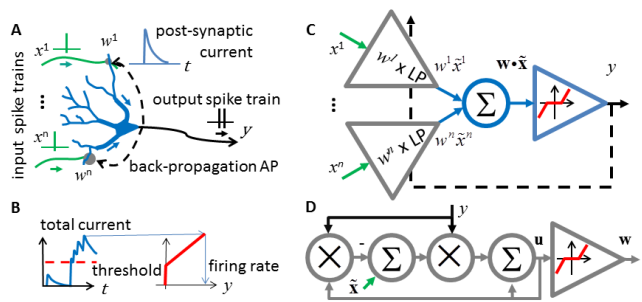

Fig. 1. A neuron as a signal processing device. *A*) A simplified view of a biological neuron. A neuron receives spike trains from presynaptic neurons. Each spike train is low-pass-filtered, or leaky integrated, and weighted by synapses. Synaptic weights are modified by Hebbian-like rules according to the correlation between the inputs and the output communicated by backpropagating action potentials (AP, dashed lines). *B*) Synaptic currents are summed together and the total current is soft-thresholded by the spike-generation mechanism. *C*) A signal processing view of a neuron. Output firing rate is given by the soft-thresholded sum of weighted LP-filtered pre-synaptic inputs. *D*) Synaptic weights are updated based on the output firing rate and LP-filtered inputs.

Whereas our knowledge of neurophysiology is much more extensive than these basic facts, the function of a neuron from a computational perspective remains a mystery. Previous attempts have been made to assign computational function to several aspects of neuronal physiology. For example, a leaky integrate-and-fire neuron has been derived from an optimal change detection in neuronal inputs [1], from minimization of prediction error for a linear dynamical system [2], and optimal estimation of presynaptic membrane potential [3]. Hebbian learning rules may enable principal component extraction if the output function is linear [4] or the detection of high-order correlations in synaptic inputs if the output function is non-linear [5]. However, these works address only a part of neural function, either activity dynamics or synaptic plasticity. A unified view of neuronal physiology from a computational perspective is still missing.

In this paper, we make a step towards a unified computational model of a neuron as a signal processing device. We adopt the firing rate approximation, in which spike timing is ignored and neuronal activity is quantified by the number of spikes per unit time. We postulate that a neuron's objective is to represent its high-dimensional input by the synaptic weight vector scaled by its output. Mathematically, a neuron finds a sparse rank-1 approximation to the input data matrix by minimizing a regularized squared representation error. By solving this minimization problem online we derive a computational algorithm that, amazingly, reproduces many salient physiological properties of a neuron.

In a departure from previous normative approaches that optimized performance in a statistical setting [1-3,5], we adopt a competitive online setting [6-8]. We choose this approach combining elements of game theory, machine learning and convex optimization because competitive online algorithms capture several aspects of neuronal function. Specifically, online algorithms compute output recursively at each time-step, without storing previous observations in memory, thus reflecting the real-time nature and low storage capacity of single neurons. Most interestingly, the performance of online algorithms is guaranteed not on average as in the statistical setting but for any possible input. Such performance guarantees convey robustness to neuronal operation.

This paper is organized as follows. In Section II, we introduce the cost function and give sparse rank-1 representation algorithms for both offline and online settings. In Section III we study the performance of the online algorithm both numerically and analytically. In Section IV, we demonstrate that the steps of the online algorithm parallel physiological properties of a biological neuron.

II. WEIGHTED SPARSE RANK-1 MATRIX FACTORIZATION

A. Offline setting

We start by postulating that a neuron computes a rank-1 representation of a temporally smoothed incoming data matrix. The data matrix, $\mathbf{X} \in \mathbb{R}^{M \times T}$, contains the activity of $M$ potential upstream neurons at $T$ time points, $\mathbf{X} = [\mathbf{x}_1, \mathbf{x}_2, ..., \mathbf{x}_T]$. The neuron represents $\mathbf{X}$ by an outer product of two sparse vectors: synaptic weights, $\mathbf{w} \in \mathbb{R}^M$, and outgoing activity, $\mathbf{y} \in \mathbb{R}^T$, which minimize the cost function comprising the discounted representation error and sparsity-inducing $\ell_1$-norm regularizers. Assuming that the data matrix is temporally correlated on a timescale, $\tau$, to improve signal-to-noise ratio the representation error is convolved with a smoothing kernel given by the powers of $\beta = \exp(-1/\tau), 0 \le \beta < 1$:

$$\{\mathbf{w}, \mathbf{y}\} = \arg\min_{\mathbf{w}, \mathbf{y}} \sum_{t=1}^{T} \left[ (1-\beta) \sum_{s=0} \beta^s \|\mathbf{x}_{t-s} - \mathbf{w} y_t\|_2^2 + 2\lambda_y |y_t| + 2\lambda_{w1} \|\mathbf{w}\|_1 + \lambda_{w2} \|\mathbf{w}\|_2^2 \right] =$$

$$= \arg\min_{\mathbf{w}, \mathbf{y}} \sum_{t=1}^{T} \left[ \|\tilde{\mathbf{x}}_t - \mathbf{w} y_t\|_2^2 + 2\lambda_y |y_t| + 2\lambda_{w1} \|\mathbf{w}\|_1 + \lambda_{w2} \|\mathbf{w}\|_2^2 \right]$$

where $\tilde{\mathbf{x}}_T = (1-\beta) \sum_{s=0} \beta^s \mathbf{x}_{T-s}$, LP filtered, or leaky-integrated vector of presynaptic inputs. While the offline cost above is convex in $\mathbf{w}$ or $\mathbf{y}$ separately, it is not jointly-convex. Such problem may be solved by the block-coordinate-descent or nonlinear Gauss-Seidel algorithm [9], which alternates minimization with respect to each variable [10,11]:

**Algorithm I: Offline sparse rank-1 matrix factorization**
Initialize $\mathbf{w}, \mathbf{y}$
Iterate until convergence:

$$\begin{cases} \mathbf{y} = \text{ST}\left(\tilde{\mathbf{X}}^T \cdot \mathbf{w}, \lambda_y\right) / \|\mathbf{w}\|_2^2 \\ \mathbf{w} = \text{ST}\left(\tilde{\mathbf{X}} \cdot \mathbf{y}, T\lambda_{w1}\right) / \left(\|\mathbf{y}\|_2^2 + T\lambda_{w2}\right) \end{cases}$$

where ST denotes soft thresholding, Figure 3C:

$$\text{ST}(f, \lambda) = \begin{cases} f - \lambda, & f > \lambda \\ 0, & |f| \le \lambda \\ f + \lambda, & f < -\lambda \end{cases}$$

When $f$ is a vector, ST acts on it in a component-wise manner.

While Algorithm I converges quickly and yields an offline solution to the sparse rank-1 factorization, it requires simultaneous access to the entire data matrix which may not be always possible. Simultaneous access is impossible when the data matrix is too large to be loaded into computer memory at the same time or when the data matrix is streamed, one column at a time, and a (partial) result must be computed in real time. Since neurons cannot store streaming data in memory and must compute on it in real time, their function must be analyzed in the so-called online setting.

B. Online setting

In the online setting the cumulative loss must be minimized by computing both $\mathbf{w}$ and (partially) $\mathbf{y}$ in real-time but only from the data received so far. For $T \gg \tau$ we have:

$$\{\mathbf{w}_T, y_T\} = \arg\min_{\mathbf{w}, y} \sum_{t=1}^{T-1} \left[ (1-\beta) \sum_{s=0} \beta^s \|\mathbf{x}_{t-s} - \mathbf{w} y_t\|_2^2 \right] +$$

$$+ (1-\beta) \sum_{s=0} \beta^s \|\mathbf{x}_{T-s} - \mathbf{w} y\|_2^2 + 2\lambda_y |y| + 2T\lambda_{w1} \|\mathbf{w}\|_1 + T\lambda_{w2} \|\mathbf{w}\|_2^2$$

Again, we solve this online minimization problem by alternating the minimization steps with respect to $\mathbf{w}$ and $y$. First, we fix the value of $\mathbf{w}$ and minimize the online loss with respect to $y$:

$$y_T = \arg\min_y (1-\beta) \sum_{s=0} \beta^s \|\mathbf{x}_{T-s} - \mathbf{w}_{T-1} y\|_2^2 + 2\lambda_y \|y\|_1 =$$

$$= \arg\min_y \left[ \frac{(1-\beta) \sum_{s=0} \beta^s \mathbf{x}_{T-s} \cdot \mathbf{w}_{T-1}}{\|\mathbf{w}_{T-1}\|_2^2} - y \right]^2 + \frac{2\lambda_y}{\|\mathbf{w}_{T-1}\|_2^2} \|y\|_1$$

We solve this minimization problem by leaky integrating presynaptic inputs, $\tilde{\mathbf{x}}_T = (1-\beta) \sum_{s=0} \beta^s \mathbf{x}_{T-s}$, and soft thresholding their weighted sum:

$$\begin{cases} \tilde{\mathbf{x}}_T = \beta \tilde{\mathbf{x}}_{T-1} + (1-\beta)\mathbf{x}_T \\ y_T = \mathrm{ST}(\mathbf{w}_{T-1} \cdot \tilde{\mathbf{x}}_T, \lambda_y) / \|\mathbf{w}_{T-1}\|_2^2 \end{cases}$$

These equations are nothing else but a commonly used linear-nonlinear or leaky integrate-and-fire model of a neuron, Fig. 1C. Note that by using recursion we avoided storing past data keeping this step truly online.

Second, we fix the value of $y$ and minimize the online loss with respect to $\mathbf{w}$:

$$\mathbf{w}_T = \arg\min_{\mathbf{w}} \sum_{t=1}^{T}\left[(1-\beta)\sum_{s=0}\beta^s \|\mathbf{x}_{t-s} - \mathbf{w} y_t\|_2^2 + 2\lambda_{w1}\|\mathbf{w}\|_1 + \lambda_{w2}\|\mathbf{w}\|_2^2\right] =$$

$$\arg\min_{\mathbf{w}} \left\| \frac{\sum_{t=1}^{T} y_t (1-\beta)\sum_{s=0}\beta^s \mathbf{x}_{t-s}}{\sum_{t=1}^{T} y_t^2} - \mathbf{w} \right\|_2^2 + T\frac{2\lambda_{w1}\|\mathbf{w}\|_1 + \lambda_{w2}\|\mathbf{w}\|_2^2}{\sum_{t=1}^{T} y_t^2}$$

By introducing a cumulative squared postsynaptic activity, $Y_T = \sum_{t=1}^{T} y_t^2$, we obtain a truly online solution, or a synaptic learning rule (see Fig.1D), which together with the above activity dynamics yields:

**Algorithm II: Online sparse rank-1 matrix factorization**

$$\begin{cases} \tilde{\mathbf{x}}_T = \beta \tilde{\mathbf{x}}_{T-1} + (1-\beta)\mathbf{x}_T \\ y_T = \mathrm{ST}(\mathbf{w}_{T-1} \cdot \tilde{\mathbf{x}}_T, \lambda_y) / \|\mathbf{w}_{T-1}\|_2^2 \\ Y_T = Y_{T-1} + y_T^2 \\ \mathbf{u}_T = \mathbf{u}_{T-1} + y_T(\tilde{\mathbf{x}}_T - \mathbf{u}_{T-1} y_T)/Y_T \\ \mathbf{w}_T = \mathrm{ST}(\mathbf{u}_T, T\lambda_{w1}/(Y_T + T\lambda_{w2})) \end{cases}$$

While the synaptic weight update equation is similar to the Oja rule [4], it was derived not by linearizing the cost function, like the Oja rule, but by solving the minimization problem exactly. Thus it does not contain an arbitrary learning rate and is, in this sense, parameter-free. Moreover, physiological synaptic weights, $\mathbf{w}$, are obtained by soft thresholding an internal variable, $\mathbf{u}$, a suggestion that, to our knowledge, has not been made previously.

Algorithm II is closest to subspace tracking algorithms [12-14] and the Oja rule [4] but there are several differences. First, unlike previous work, we regularize both weights and activity resulting in nonlinearities. Second, unlike [12,13] who applied discounting to $\mathbf{w}$, we apply the discounting with respect to relative time, $s$, resulting in leaky integration in the calculation of $y$. If $\mathbf{w}$ is expected to vary over time one can easily introduce discounting with respect to the absolute time, $t$. Such discounting would trade off memory longevity for adaptability.

## III. PERFORMANCE ANALYSIS OF THE ONLINE ALGORITHM

In this section, we analyze the performance of Algorithm II numerically and analytically. First, we applied Algorithm II to a standard dataset of whitened natural image patches. Pixel intensities from each image patch were reshaped into a vector and presented to the algorithm sequentially. The results of the algorithm, Fig. 2, agree with [15] and biological observations,

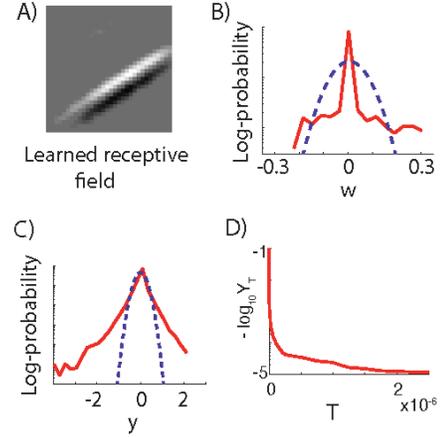

see next section.

Fig. 2. Online sparse rank-1 representation of the natural image dataset using Algorithm II ($\lambda_{w1} = 0.002$; $\lambda_y = 0.4$; $\beta = \exp(-1/10)$). The algorithm is presented with 50000, 32x32 pre-whitened natural image patches. Each image is presented 50 time steps. A) Learned weights form a Gabor-like receptive field. B) Learned weight distribution (solid red line) is heavy-tailed. Dashed blue line is a Gaussian with identical mean and variance. C) Learned weights lead to sparse firing with heavy-tailed statistics (solid red line). Dashed blue line is a Gaussian with identical mean and variance. This plot was obtained by presenting the neuron with the same image patches but synaptic weights were frozen. D) Learning rate drops over time.

Second, we would like to demonstrate that Algorithm II, despite operating online, performs, asymptotically in the large $T$ limit, no worse than Algorithm I. Analysis is complicated by the fact that even in the offline setting the problem is not convex and optimal solution is not guaranteed. Because of this, following [16] we focus on the algorithm's ability to learn synaptic weights, $\mathbf{w}$. For that purpose we consider a repeated game [6-8], with the following rules for each time step, $t=1\ldots T$:

1. Algorithm selects value $\hat{\mathbf{w}}_t$

2. World reveals $\mathbf{x}_t$ and $y_t$

3. Algorithm suffers loss:

$$l_t(\hat{\mathbf{w}}_t) = \|\tilde{\mathbf{x}}_t - \hat{\mathbf{w}}_t y_t\|_2^2 + 2\lambda_{w1}\|\hat{\mathbf{w}}_t\|_1 + \lambda_{w2}\|\hat{\mathbf{w}}_t\|_2^2$$

The regret of the online algorithm is its cumulative loss relative to the best possible offline solution:

$$R(\hat{\mathbf{w}}_1, \ldots, \hat{\mathbf{w}}_T) = \sum_{t=1}^{T} l_t(\hat{\mathbf{w}}_t) - \min_{\mathbf{w}} \sum_{t=1}^{T} l_t(\mathbf{w})$$

The best offline solution can be found by applying the second step of Algorithm I (no iteration required). The online algorithm effectively solves the optimization problem, $\min_{\mathbf{w}} \sum_{s=1}^{t-1} l_s(\mathbf{w})$, which is equivalent to an offline problem because all needed information is available. Such online strategy is called "follow the leader" (FTL) [6-8] and has a provable regret bound.

**Theorem**: Logarithmic bound on the regret of FTL. Let $D$ be such that $y_t \|\tilde{\mathbf{x}}_t - \hat{\mathbf{w}}_t\|_2 \leq D$ for all $t$, and $d$ be such that $\|\hat{\mathbf{w}}_t\|_2 \leq d$ for all $t$, then:

$$R(\hat{w}_1,...,\hat{w}_T) \leq 16(D + \lambda_{w1} + \lambda_{w2}d)^2 (1+\log T)/\lambda_{w2}$$

**Sketch of proof:** Following [17, Theorem 3.1], the upper bound on the regret for a convex twice differentiable objective function is: $4B^2(1+\log T)/C$, where $B$ is Lipschitz constant and $C$ is the lower limit on Hessian eigenvalues. The proof can be modified for non-differentiable function, which is strongly convex with parameter $C$. Substituting $B = 2(D + \lambda_{w1} + \lambda_{w2}d)$, $C = \lambda_{w2}$, we obtain the regret bound.

According to the above theorem, the cumulative loss of the online algorithm relative to the offline is no worse than $\log T$, meaning that the difference in loss per round for online relative to offline goes to zero in the large $T$ limit. In this sense, the online algorithm performs asymptotically no worse than offline.

## IV. COMPARISON OF THE ONLINE RANK-1 FACTORIZATION ALGORITHM WITH OBSERVATIONS ON BIOLOGICAL NEURONS

In this section, we demonstrate that the steps of the online algorithm derived in the previous section, as well as the statistics of its output, are amazingly similar to the salient physiological properties of biological neurons.

*1) Synaptically weighted summation of pre-synaptic activity.* Neurons are known to sum weighted presynaptic activity by combining the corresponding postsynaptic currents according to Kirchhoff's current law, Fig.1. While theoretical models of a neuron often postulate weighted summation, we obtain it from minimizing a principled loss function.

*2) Leaky integration.* Discounting of errors in representing past inputs, a common requirement in signal processing, results in leaky integration of inputs to a neuron, Figs. 1A,3A. The experimentally observed signature of leaky integration is a time course of a postsynaptic current in response to a presynaptic spike, Fig. 3B. Our framework suggests that the time constant of the decay should depend on the correlation time scale of the input and its SNR.

*3) Nonlinear output function.* Inclusion of a sparsity inducing regularizer, such as an $l_1$-norm, of activity into the loss function leads to a nonlinear operation on the total current, such as soft thresholding, Figs. 1B,3C. A nonlinear firing rate vs. current curve experimentaly observed in many neurons, Fig. 3D, mimics one side of the soft thresholding function, Fig. 3C. Therefore, a pair of neurons, like ON-OFF cells in the retina, can together implement a two-sided soft-threshold function.

*4) NonGaussian distribution of neuronal activity.* As a consequence of soft thresholding derived from $l_1$-norm regularization the distribution of output activity in our model is nonGaussian, sparse and heavy-tailed, Fig. 2C. Such distribution of neuronal firing rates has been observed experimentally, Fig. 4A.

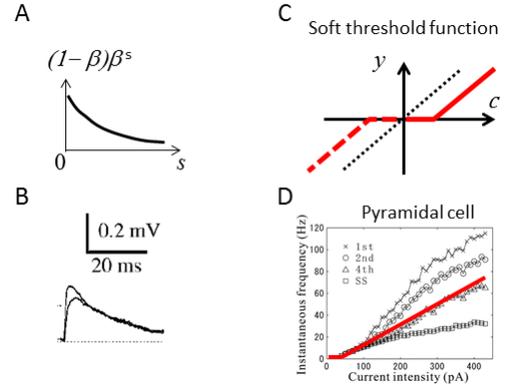

Fig. 3. Leaky integration and soft thresholding in the online algorithm and biological neurons. *A*) Leaky integration specified by the discounting factor. *B*) Experimentally measured decay of a postsynaptic potential in response to a presynpatic spike is similar to leaky integration [18]. *C*) Because of the $l_1$-norm regularizer sparse factorization algorithms apply soft threshold function to the weighted sum of leaky integrated inputs. *D*) Experimentally measured firing rate vs. injected current curve [19] is similar to the nonnegative half of the soft threshold function.

*5) Synaptic weight update reflects the correlation between pre- and postsynaptic activity (Hebb postulate) with an activity dependent learning rate.* Unlike the arbitrary learning rate in the Oja rule, our learning rate is the inverse of the cumulative squared postsynaptic activity. This result parallels the experimentally reported decay of synaptic plasticity, known as LTP, with age in an activity-dependent manner [21-23].

*6) Soft thresholding step in synaptic weight update predicts the existence of silent synapses.* If the cumulative correlation between pre- and postsynaptic activity exists (non-zero $u$) but does not reach the threshold specified by the regularization coefficient then the physiological synaptic weight, $w$, should be zero. Such silent synapses, i.e. morphologically defined synapses without a physiological synaptic weight, have been observed experimentally [24,25].

*7) NonGaussian distribution of synaptic weights.* As a consequence of soft thresholding derived from $l_1$-norm regularization the distribution of synaptic weights must be sparse and heavy-tailed. Ineed, both physiologically [20] and anatomically [26,27] measured synaptic weights follow a sparse heavy-tailed distribution, Fig. 4B.

*8) Learning Gabor features from the natural scene ensemble.* When presented with a dataset of whitened patches from natural images this algorithm learns a Gabor feature similar to the receptive field of neurons in the primary visual cortex [28].

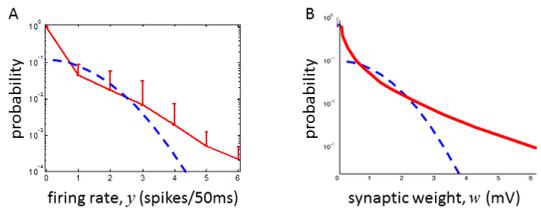

Fig. 4. Experimentally observed nonGaussian distributions. Red: distribution of firing rates (*A*) (Courtesy of J. Magee) and synaptic weights (*B*) [20] are nonGaussian, sparse and heavy-tailed. Dashed blue: a Gaussian distribution shown for comparison.

## V. CONCLUSION

We postulate that the computational function of a neuron is to represent a streaming data matrix of presynaptic activity by an outer product of its synaptic weights and the outgoing activity as a function of time. We derive an online algorithm that computes such representation and demonstrate that it reproduces many physiological properties of a neuron. When trained on natural images the algorithm learns Gabor-like features as observed in the primary visual cortex. The performance of the online algorithm is asymptotically no worse than that of the offline one conveying a degree of robustness. Thus, we make a step towards a unified computational model of a neuron that should help model neuronal networks without necessarily measuring all the biophysical parameters and help design neuromorphic electronics.

## ACKNOWLEDGMENT

We are grateful to Jeff Magee and Elad Hazan for helpful discussions.